\documentclass[twocolumn,trackchanges]{aastex631}

\usepackage{mathptmx}
\usepackage[T1]{fontenc}
\usepackage{ae,aecompl}
\usepackage{graphicx}	
\usepackage{amsmath}	
\usepackage{amssymb}	

\begin{document}

\title{Gas Phase Ions in Protoplanetary Disks from Collisions of Solids}

\author[0009-0003-9097-1934]{Jakob Penner}
\affiliation{University of Duisburg-Essen, Faculty of Physics,
Lotharstr. 1-21, 47057 Duisburg, Germany}

\author[0000-0002-7962-4961]{Gerhard Wurm}
\affiliation{University of Duisburg-Essen, Faculty of Physics,
Lotharstr. 1-21, 47057 Duisburg, Germany}

\author[0000-0003-4468-4937]{Jens Teiser}
\affiliation{University of Duisburg-Essen, Faculty of Physics,
Lotharstr. 1-21, 47057 Duisburg, Germany}

\begin{abstract}
Ionization is important for magnetohydrodynamics and chemistry in protoplanetary disks but known ionization sources are often weak along the midplane. We present, for the first time, data from a laboratory experiment, where we measure ions from colliding mm-basalt grains emitted into the surrounding gas phase. This positive detection implies that very basic collisions in early phases of planet formation are sources of ionization. The midplane of protoplanetary disks might be ionized despite the lack of intense radiation sources. 
\end{abstract}


\section*{Introduction}
Ionization of protoplanetary disks affects several problems related to planet formation, mostly connected to magnetic interactions. This includes accretion by turbulence generated by instabilities like the MRI \citep{ Balbus1991, Gammie1996, Flock2012, Delage2021, Delage2023}, vertical shear instabilities in magnetized disks \citep{Cui2021}, accretion going along with magnetically launched winds \citep{Turner2014}, or disk fragmentation by a dynamo \citep{Deng2021}. In turn, these disk conditions then allow or do not allow concentration and growth of solids to become seeds of planets, e.g. by setting collision speeds \citep{Ormel2007,Yang2018, Gong2021, Mori2021}

Especially magnetic interactions depend on the ionization degree of the disk. While only small ionization rates are needed for most of the applications, the known ionization sources do provide only small ionization rates - at least in the dense midplane. Therefore, the balance is critical.

So far, ionization models have to invoke one of the well known ionization sources. These are thermal ionization in hot regions of the disk, stellar x-rays impinging the surface, cosmic rays entering somewhat deeper, and radioactive decay of short lived radioisotopes \citep{Armitage2011, Ercolano2013, Glassgold2017, Umebayashi2009, Cleeves2013, Johansen2018}.

Very recently though, another ionization mechanism for protoplanetary disks was proposed by \cite{Wurm2022} based on findings from observing collisions of solid grains by \cite{Jungmann2021b}. \cite{Jungmann2021b} found - somewhat unexpected or unnoticed before - that grains not only transfer charge among themselves in collisions by tribo-charging but also loose some charges during a collision. Obviously, this charge has to become entrained into the surrounding gas as charge cannot really be lost. As collisions come naturally in the midplane of protoplanetary disks, planet formation - in a way - might come with its very own special ionizer. 

In more detail, dust grains initially grow in collisions by hit-and-stick in early phases of planet formation. However, there will be collisional barriers opposing further growth, eventually  \citep{Wurm2021nat}. 
An early one to encounter is the bouncing barrier where aggregation gets stalled at mm-size \citep{Zsom2010, Kelling2014, Kruss2016,Arakawa2023}.
This is were tribocharging and ionization come into play.
It was shown in recent years that tribo-charging going along with collisions might eventually support further growth into larger aggregates \citep{Steinpilz2019}. In any case, charging is an important part in collisions of particles in protoplanetary disks.

So far, ionization of the disk's gas phase and early collisional evolution of pre-planetary bodies are quite distinct topics.
\cite{Wurm2022} estimated though that the ionization rate due to grain collisions in the disk midplane can be larger than other ionization sources in protoplanetary disks. 
While \cite{Jungmann2021b} measured charge balances on the colliding macroscopic grains, ions were not detected directly. Here, we do detect generated gas phase ions. Our goal is to show by different means than charge measurements on grains that the ionization process of the ambient gas proposed is real.

The rationale behind the experiments which are reported is therefore as follows. Collisions of grains provide ions that become entrained in the surrounding gas. If they should be important for protoplanetary disks, these ions should become free ions, i.e. no longer tied to their parent source grains. If so, it should be possible to transport some of these ions by some gas flow to some other place. There, some device should be capable of detecting them. This is the experiment we did set up and which is described below.

\section{Experiments}

\begin{figure}
    \centering
    \includegraphics[width= \columnwidth]{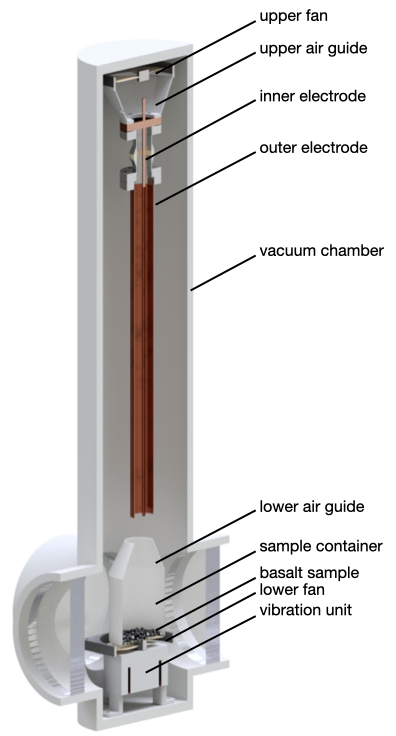}
    \caption{Schematics of the experiment; Gas is pumped by two fans through a vibrated particle bed and through a tube capacitor. Gas phase ions which are generated by collisions between grains are detected by the capacitor. The setup is placed within a vacuum chamber.}
    \label{fig:setup}
\end{figure}

Fig. \ref{fig:setup} shows the setup of the experiment. 
About 34 g of basalt particles of $800 \, .. \, 1000 \, \rm \mu m$ in size (diameter) are placed within a plastic container. Measuring the mass of 21 spheres to 
26 mg, one particle has an average mass of 
1.2 mg. The container with a diameter of 9 cm therefore contains about 
28000 particles, which is about 4 layers of grains. The container bottom consists of a steel mesh at the bottom which supports the grains but allows a gas flow through the sample.  
Gas flow is provided by a fan below the sample which pumps gas through the sample. 
The container and fan are installed on a voice coil that allows the sample to be vibrated which induces particle collisions and collisional charging.

About 20 cm above the bottom of the sample container, a cylindrical capacitor made of copper is situated with a center electrode and an outer cylinder with an inner diameter of 3,9 cm. A plastic guide for the gas flow is installed between sample container and capacitor.
On top of this Gerdien-tube like setup, another fan pumps air upwards, supporting an upward flow of gas through the sample and along the electrodes.
The whole setup is placed within a vacuum chamber with an inner diameter of 21,3 cm to allow pressure dependent measurements down to below 1 mbar.

\begin{figure}
    \centering
    \includegraphics[width= \columnwidth]{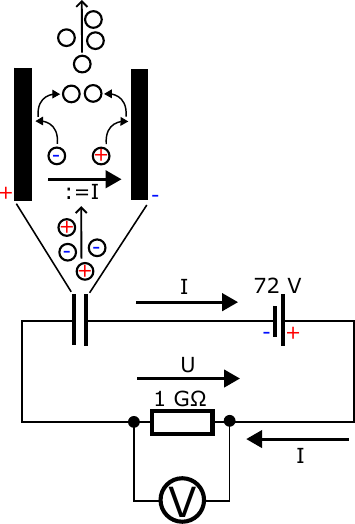}
    \caption{{Circuit diagram to visualize the basic measurement. The capacitor allows a current related to the ion pairs per time arriving. This current is driven by a DC voltage as a series of batteries. Both polarities of ions are needed. The current is detected as voltage drop on a resistor with high impedance.}}
    \label{fig:circuit}
\end{figure}

{As seen in fig. \ref{fig:circuit}, the capacitor is in line with a power supply of 72 V made from 8 x 9 V batteries and in line with an ohmic resistor of $1\,\rm G \Omega$. While we use a capacitor setting we actually do not measure the charging or discharging of the capacitor but it is only used to limit the current to the ion pairs / per second arriving. If ions (pairs of both polarities) are collected by the electrodes, this translates in a current and voltage drop along the resistor which is measured by a Keithley 2182 Nanovoltmeter.}

The particle container is vibrated with 100 Hz. For reference, also measurements without sample were carried out.

\section{Results}

\begin{figure}
    \centering
    \includegraphics[width= \columnwidth]{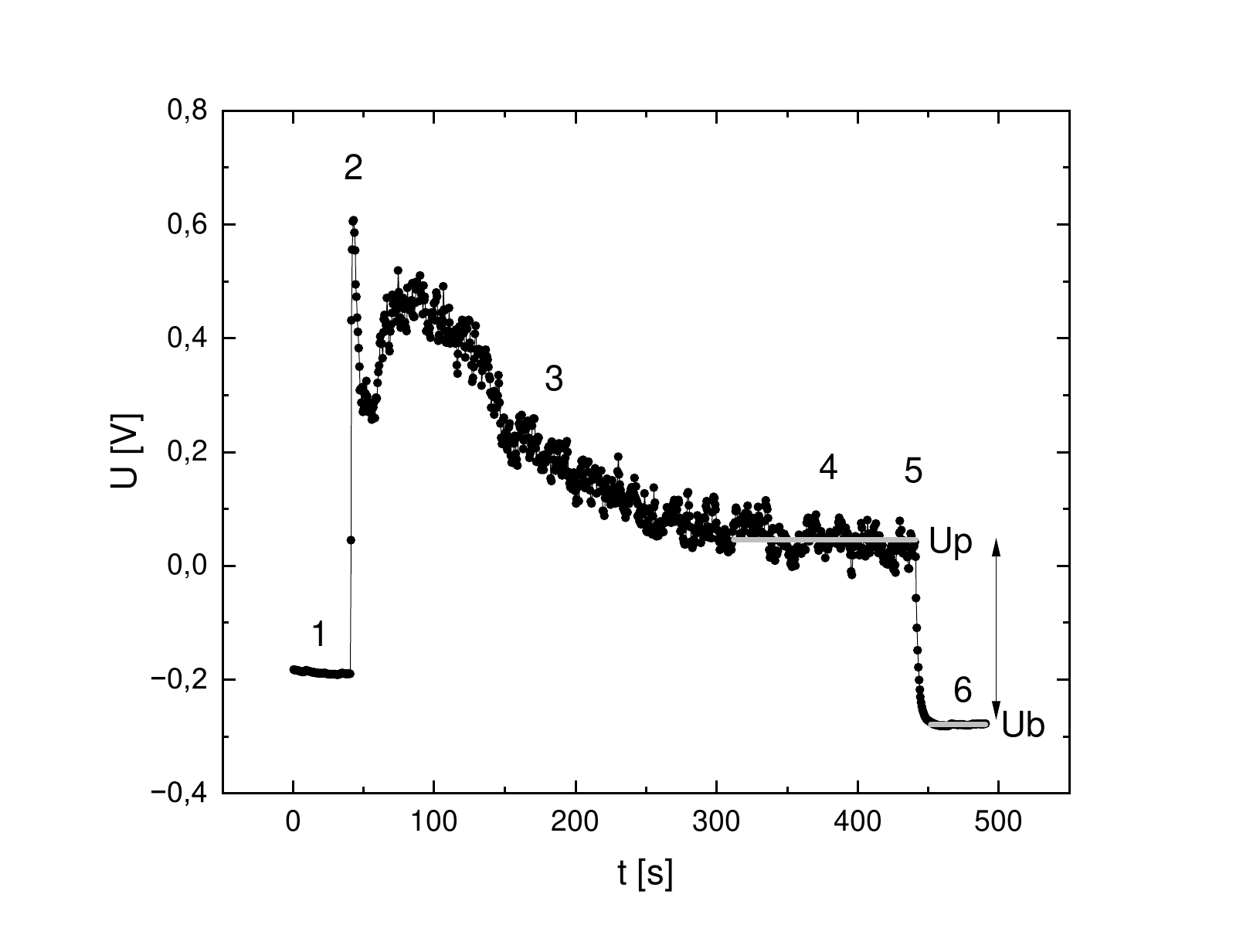}
    \caption{Example measurement at 1 mbar; Voltage $U$ is proportional to the detected ion rate; Numbered features refer to 1) background level before vibrations start; 2) vibrations start; 3) initial maximum and decline; 4) equilibrium value; 5) vibrations stop; 6) background level.}
    \label{fig:dataone}
\end{figure}

No current was measured along the electrodes if the experiment was run without granular sample but ions were detected if the sample was included. This verifies that we do not measure electrical signals generated by the active experiment, i.e. by the fans or shaking voice coil. 
Fig. \ref{fig:dataone} shows an example of the basic measurement where the voltage $U$  refers to the voltage drop on the resistor as the ions provide the charge carriers for a current within the measurement circuit. The measured voltage is therefore proportional to the ion rate detected. The data shown are for an ambient pressure of 1 mbar, which is on the higher end of pressures expected for protoplanetary disks \citep{Wood2000}. Marked within this example are all significant times and features of the data. 

\subsection{Signal features}

Seen in all data is a peak at the onset of vibrations (marked 2 in fig. \ref{fig:dataone}) which varies strongly in amplitude though. It strongly depends on the history of the measurement sequences. It is always high at normal atmospheric pressure, is high at the first low pressure ((sub)-mbar range) measurement but only small if more low pressure measurements follow only minutes to hours later. Obviously, this is indicative of some electrostatic fields which are set up by the vibrations and do or do not relax after the experiment, as electrostatic fields influence the ion flow.

There is some highly variable kind of signal (marked 3) following the initial peak, which we currently attribute to various electrostatic fields which are generated by the colliding grains and which deflect the ion flow. This is discussed below in somewhat more detail. In any case, the signal, independent of the history of experiments, convergence to an equilibrium value marked as (4), eventually. Due to its constant nature, this feature is well suited for a systematic analysis of the effect of parameter variations and we take the difference between the value at this plateau at the end of vibrations (5), $U_p$, and the background level, $U_b$, marked as (6) in fig. \ref{fig:dataone}. 

{We note that the constant signal implies that both polarities of ions are generated continuously. This does not rule out a bias towards one polarity ions which might engulf particles in an ion cloud. However, if only ions of one polarity reached the capacitor, the capacitor would rapidly charge which resulting in a very short pulse, which is not detectable here.  Without ion pairs entering, the capacitor just acts as an isolator. The positive signals do therefore not imply an ion polarity but are just the direction, the voltage is measured along the resistor. The slightly negative values of the base line are a bias induced by the high sensitivity electrometer.}

For this first work of its kind, we varied the ambient pressure as parameter to test the hypothesis that ionization might work at pressures of protoplanetary disks. 
{We use the equilibrium values of the signals to yield some qualitative comparisons but clearly note that the quantitative deduction of ion generation rates from this as given below only yields plausible values in agreement to earlier work \citep{Jungmann2021b}. Explanations of the initial time evolution of the signal features, as given below, are currently speculative.
However, we also note that the mere existence of ion detection apart from any equilibrium value shows that collisions of grains constantly produce ion pairs that become entrained into the gas flow, which - to the knowledge of the authors - has never been seen before.}

\subsection{Pressure dependent ionization}

\begin{figure}
    \centering
   \includegraphics[width= \columnwidth]{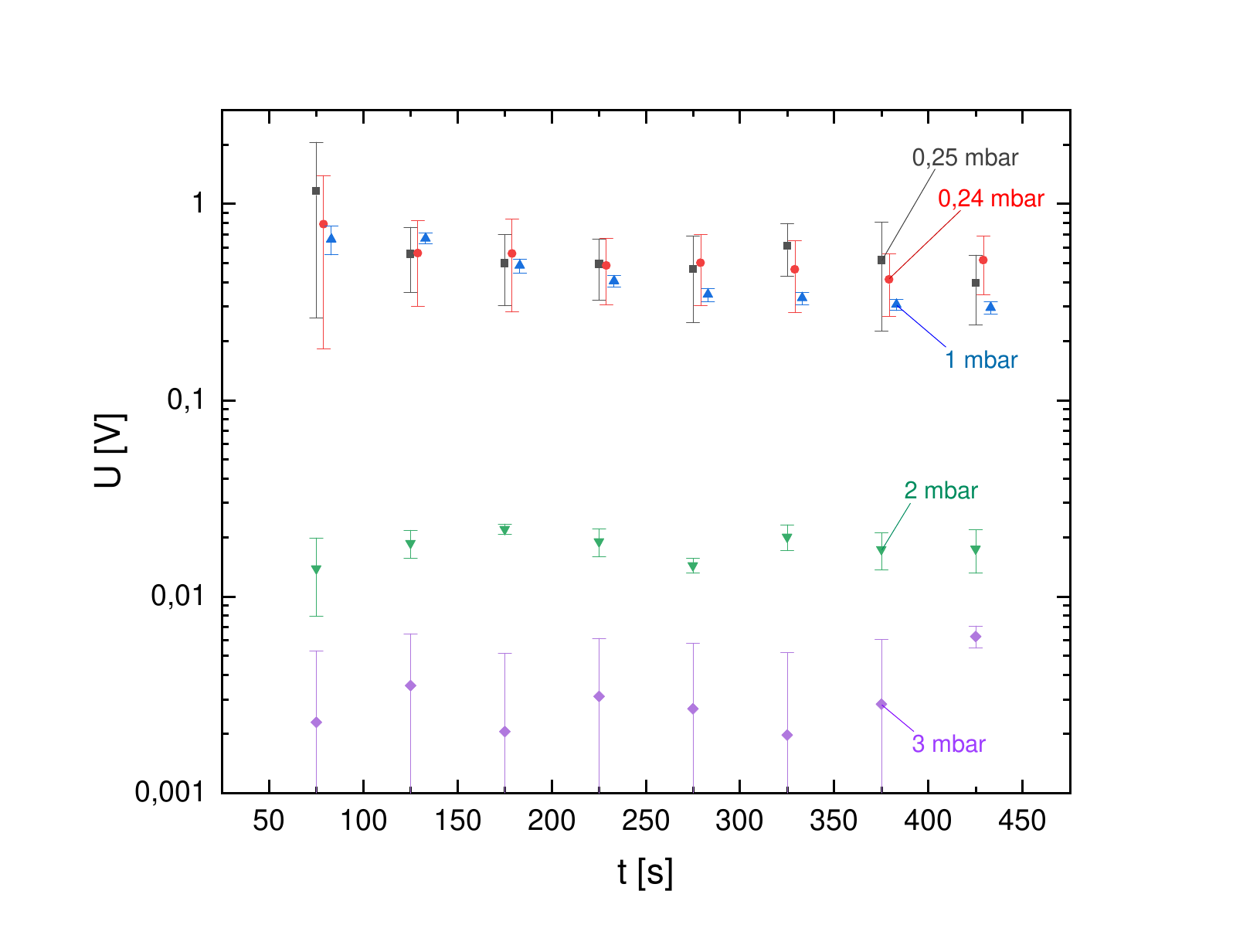}
    \caption{Equilibrium voltage for different pressures in the (sub)-mbar range. Data is averaged over 50 s and related to the background level or $U = U_p - U_b$. Error bars are standard deviations and reflect the strength of fluctuations. Fluctuations for the 3 mbar measurement reach down to zero but are much smaller absolutely compared to the fluctuations at sub-mbar pressure (note the log scale).}
    \label{fig:pressure}
\end{figure}

\cite{Becker2022} found that tribocharging of grains essentially works at all pressure ranges relevant for protoplanetary disks, i.e. down to $\rm 10^{-8} \, mbar$.  Our individual measurements for different ambient pressures are shown in fig. \ref{fig:pressure}. This plot only focuses on the data during vibration. The data are binned and related to the background value, i.e. show $U_p - U_b$.
The error bars give the standard deviations. The value at 3 mbar is still distinctive from the background. Measurements at higher pressure do not result in an equilibrium signal level that can be resolved with the current setup. The lowest pressures come with the highest signal but also with the largest fluctuations.  Measurements at a pressure below 0.2 mbar are not within the range of the current setup and require a different design of the setup as the transport capabilities of the fans decrease in efficiency.

There is a clear tendency that starting at a few mbar the equilibrium ion flux is increasing toward lower pressure. 
Care has to be taken though. Not shown here, there are large peaks (2 in fig. \ref{fig:dataone}) at the beginning of vibrations. These also occur at normal pressure, even though there is a lack of a measurable convergence value afterwards. 
A lack in equilibrium value therefore does not necessarily imply that no ions are produced at high pressure. In fact, the results of \cite{Jungmann2021b} were taken at normal pressure. Therefore, the interpretation might be more complex. Discharge processes and electrostatic deflecting fields as well as the gas flow might change all strongly for the different pressure ranges. Keeping all these currently unknown details in mind, the low pressure range between 0.25 and 3 mbar suggests that lower pressures are at least good for producing ions. Protoplanetary disks might therefore provide environments where grain collision ionization might work well.

\section{Discussion}

\subsection{Charge sinks and the initial peak}

\begin{figure}
    \centering
   \includegraphics[width= \columnwidth]{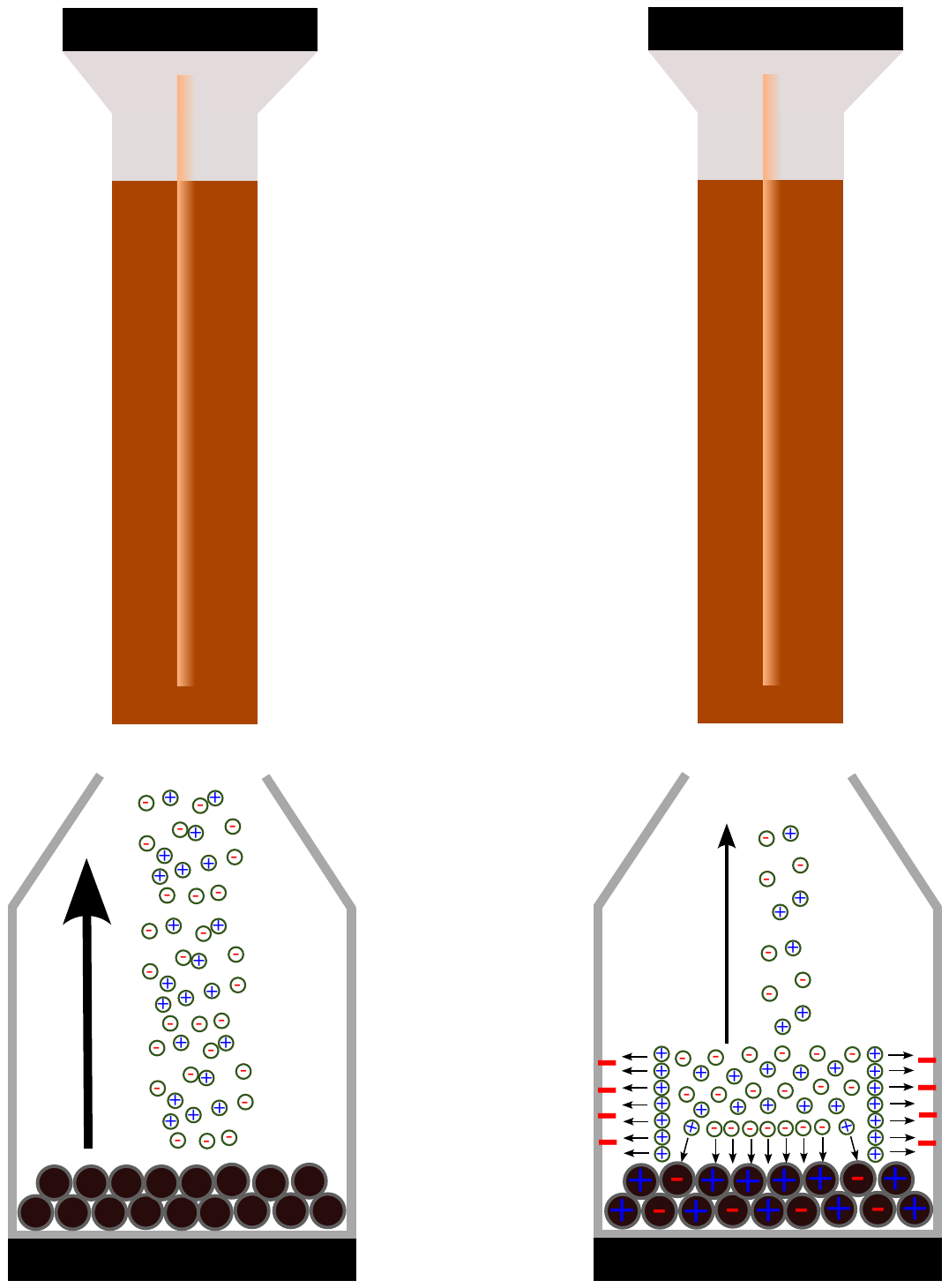}
    \caption{A sketch of the ion transport at the start of the experiment (left) and in equilibrium (right); As charges and electric fields build up on walls and grains, only part of the generated gas-phase ions manage to leave at later times while a large number of charges is trapped and recombines locally.}
    \label{fig:reduction}
\end{figure}

The experiments prove that collisions produce free ions (ion pairs) which become entrained into a gas flow. However, the electrostatic fields that build up have a large impact on the ion flow. One component of such an electrostatic field will be generated by the particle bed itself. With increasing charge on the grains, ions become more prone to recombine on other grains close by. 
A second component of an electrostatic field originates from the material difference between sample particles (basalt), container walls (plastic), and supporting mesh (metal). Different material tribocharging usually leads to one material being charged with a different polarity than the other, e.g. according to a triboelectric series \citep{Shaw1917}. 
These fields will also deflect ions from their upward flow. Such kinds of fields were e.g. observed as visible corona discharges in experiments by \cite{Schoenau2021}.
This idea of deflecting fields which decrease the upward ion flux is sketched in fig. \ref{fig:reduction}.

In view of these fields, the equilibrium level in fig. \ref{fig:dataone} (5) is only a lower limit of the ions that are produced. The initial peak (2) might be closer to an unbiased flux from the sample which would be more relevant for the dilute setting of protoplanetary disks. With potential peaks up to several V, there is a large reduction factor of at least 10 for the constant ion flow due to the electrostatic shielding of the ion flux.

{\subsection{Ion species}}

{There are at least two kinds of ions produced as two polarities are necessary to generate a constant current. So there are positive and negative ions. In view of the observed pressure dependence, gaseous breakdown is a potential source. \cite{Cruise2023} find that gaseous breakdown might also limit charges on grains at higher, i.e. normal atmospheric pressure. And also \cite{Jungmann2021b} found the charge imbalance under normal pressure. If gaseous breakdown generates the ions then all gas phase molecules likely belong to the initial ion inventory. That means e.g. $\rm O_2^+$ or $\rm H_3O^+$  on the positive side \citep{Skalny2006}. If we consider the discharge to form similar ions as an atmospheric corona discharge, quite a number of negative ions might result from $\rm O^-, O_2^-, O_3^-$ over $\rm CO_3^-$ to $\rm NO_3^-$ likely embedded in some water cluster \citep{Skalny2006, Jiang2018}. }

{Gaseous breakdown might not be responsible for all ions though.  
As water is often considered as a possible charge agent for tribocharging \citep{Jungmann2022x, Lee2018},  $\rm H_3O+$ and $\rm OH^-$ are relevant candidates as prime ions to be shed from the surfaces. This mechanism might work down to one monolayer of water at very low pressure. At least, charging of the grains is still observable \citep{Becker2022}. To which pressure gas phase ions are produced is an open question though. As the formation scenarios are different from atmospheric (1 bar) discharges it is currently speculative what the most abundant ion species might be. This is especially true in view of protoplanetary disks with a quite distinct atmosphere dominated by hydrogen and helium.}

\subsection{Ion rate estimate}

The equilibrium voltage measured gives a current which directly measures the rate of ion pairs detected. E.g. at $\Delta U = 1\, \rm V$ on a $R = 1 \,\rm G \Omega$ resistor, this would be $I = 1\, \rm nC/s$.
This measured current can be translated into an ionization charge per collision $Q$ as follows:

\begin{equation}
\label{gleichung}
    Q = \frac{I  }{N \cdot \nu } f_{exp}
\end{equation}

where $N$ is the number of grains and $\nu$ is the collision frequency.
The current, the particle number and the collision frequency only set a very lower limit though as detailed in the previous subsection. A correction factor $f_{exp}$ has to be included that accounts for the experimental setting and all deviations from the charge liberated in an individual collision within a protoplanetary disk. If the peak upon initial start of the vibrations is indicative for a field free ion transfer before the fields build up, the correction factor might be at least on the order of 10.

The quantities $N$ and $\nu$ are set by the experiment.
There might be a reduction to an efficient $N$ due to particles being too close together to each other in a particle bed. This is connected to the charging / discharging mechanisms.
It is currently still unknown, why grains "loose" charge into the surroundings. One reasonable assumption is that grains are themselves charged so highly that upon approach or retraction the conditions for a gaseous breakdown would be fulfilled. Then the gas would be ionized as part of a Townsend avalanche \citep{Krauss2003, Harper2016, Matsuyama2018, Wurm2019, Schoenau2021}.
In any case, if gas breakdown drives ionization, the condition for breakdown might not be reached between close grains but only for few grains hopping high enough, separating from the others \citep{Schoenau2021}.
We therefore assume here that only the top layer of vibrated grains emits measurable charges. These are about $N = 10.000$ grains then.
The collision frequency at maximum is the vibration frequency of $100 \rm Hz$. However, the top layer of particles moves freely between kicks of the actuator and particles move up to about 5 cm. Then the free fall time is about 0.2 s. Assuming this, the collision frequency is $\nu = 5 \rm Hz$. 

We cannot really constrain $f_{exp}$ well but nevertheless give a first plausibility estimate here. 
We assume a factor 10 for field deflection of the walls, a second factor 10 for fields between grains within the bed collecting charges, a third factor 10 reduction for suppressed ionization of too close grains and maybe a fourth factor 10 for recombination related to a high density of ions within the flow. This gives a total of $f_{exp} = 10^4$. 
The charge generated per collision is then (eq. \ref{gleichung}) $Q \sim 200 \,\rm pC$. This is on the order of magnitude determined from the charge balance on individual grains by \cite{Jungmann2021b}. 
We clearly note that this is only a plausibility check. The true value might be lower but is likely larger if we consider the results by \cite{Jungmann2021b} as lower limit.

\section{Conclusion}

We did set up a first experiment to prove that collisions of (sub)-mm-size grains ionize the surrounding gas by directly detecting gas ions of a sample of colliding grains for the first time. The number of charges produced per grain collision is not well constrained but estimates are consistent with observations of charge losses observed on colliding grains \citep{Jungmann2021b}. The results leave no doubt that collisions of grains ionize the gas and confirm the idea proposed by \cite{Wurm2022} that collisions are an efficient ionization source in protoplanetary disks.

\section{Acknowledgments}
The project is funded by the Deutsche Forschungsgemeinschaft (DFG, German Research Foundation) – 521602700. This project is also supported by DLR Space Administration with funds provided by the Federal Ministry for Economic Affairs and Climate Action (BMWK) under grant number DLR 50 WM 2142.
We thank the anonymous referee for reviewing our manuscript.

\section{Data availability}
Data will be made available upon reasonable request.

\bibliography{bib}

\begin{thebibliography}{}
\expandafter\ifx\csname natexlab\endcsname\relax\def\natexlab#1{#1}\fi
\providecommand{\url}[1]{\href{#1}{#1}}
\providecommand{\dodoi}[1]{doi:~\href{http://doi.org/#1}{\nolinkurl{#1}}}
\providecommand{\doeprint}[1]{\href{http://ascl.net/#1}{\nolinkurl{http://ascl.net/#1}}}
\providecommand{\doarXiv}[1]{\href{https://arxiv.org/abs/#1}{\nolinkurl{https://arxiv.org/abs/#1}}}

\bibitem[{{Arakawa} {et~al.}(2023){Arakawa}, {Okuzumi}, {Tatsuuma}, {Tanaka},
  {Kokubo}, {Nishiura}, {Furuichi}, \& {Nakamoto}}]{Arakawa2023}
{Arakawa}, S., {Okuzumi}, S., {Tatsuuma}, M., {et~al.} 2023, \apjl, 951, L16,
  \dodoi{10.3847/2041-8213/acdb5f}

\bibitem[{{Armitage}(2011)}]{Armitage2011}
{Armitage}, P.~J. 2011, \araa, 49, 195,
  \dodoi{10.1146/annurev-astro-081710-102521}

\bibitem[{{Balbus} \& {Hawley}(1991)}]{Balbus1991}
{Balbus}, S.~A., \& {Hawley}, J.~F. 1991, \apj, 376, 214,
  \dodoi{10.1086/170270}

\bibitem[{{Becker} {et~al.}(2022){Becker}, {Steinpilz}, {Teiser}, \&
  {Wurm}}]{Becker2022}
{Becker}, T., {Steinpilz}, T., {Teiser}, J., \& {Wurm}, G. 2022, \mnras,
  \dodoi{10.1093/mnras/stac1320}

\bibitem[{{Cleeves} {et~al.}(2013){Cleeves}, {Adams}, {Bergin}, \&
  {Visser}}]{Cleeves2013}
{Cleeves}, L.~I., {Adams}, F.~C., {Bergin}, E.~A., \& {Visser}, R. 2013, \apj,
  777, 28, \dodoi{10.1088/0004-637X/777/1/28}

\bibitem[{{Cruise} {et~al.}(2023){Cruise}, {Starr}, {Hadler}, \&
  {Cilliers}}]{Cruise2023}
{Cruise}, R.~D., {Starr}, S.~O., {Hadler}, K., \& {Cilliers}, J.~J. 2023,
  Scientific Reports, 13, 15178, \dodoi{10.1038/s41598-023-42265-0}

\bibitem[{Cui \& Lin(2021)}]{Cui2021}
Cui, C., \& Lin, M.-K. 2021, On the Vertical Shear Instability in Magnetized
  Protoplanetary Disks.
\newblock \doarXiv{2105.11151}

\bibitem[{{Delage} {et~al.}(2023){Delage}, {G{\'a}rate}, {Okuzumi}, {Yang},
  {Pinilla}, {Flock}, {Stammler}, \& {Birnstiel}}]{Delage2023}
{Delage}, T.~N., {G{\'a}rate}, M., {Okuzumi}, S., {et~al.} 2023, arXiv
  e-prints, arXiv:2303.15675, \dodoi{10.48550/arXiv.2303.15675}

\bibitem[{{Delage} {et~al.}(2021){Delage}, {Okuzumi}, {Flock}, {Pinilla}, \&
  {Dzyurkevich}}]{Delage2021}
{Delage}, T.~N., {Okuzumi}, S., {Flock}, M., {Pinilla}, P., \& {Dzyurkevich},
  N. 2021, arXiv e-prints, arXiv:2110.05639.
\newblock \doarXiv{2110.05639}

\bibitem[{{Deng} {et~al.}(2021){Deng}, {Mayer}, \& {Helled}}]{Deng2021}
{Deng}, H., {Mayer}, L., \& {Helled}, R. 2021, Nature Astronomy, 5, 440,
  \dodoi{10.1038/s41550-020-01297-6}

\bibitem[{{Ercolano} \& {Glassgold}(2013)}]{Ercolano2013}
{Ercolano}, B., \& {Glassgold}, A.~E. 2013, \mnras, 436, 3446,
  \dodoi{10.1093/mnras/stt1826}

\bibitem[{{Flock} {et~al.}(2012){Flock}, {Henning}, \& {Klahr}}]{Flock2012}
{Flock}, M., {Henning}, T., \& {Klahr}, H. 2012, \apj, 761, 95,
  \dodoi{10.1088/0004-637X/761/2/95}

\bibitem[{{Gammie}(1996)}]{Gammie1996}
{Gammie}, C.~F. 1996, \apj, 457, 355, \dodoi{10.1086/176735}

\bibitem[{{Glassgold} {et~al.}(2017){Glassgold}, {Lizano}, \&
  {Galli}}]{Glassgold2017}
{Glassgold}, A.~E., {Lizano}, S., \& {Galli}, D. 2017, \mnras, 472, 2447,
  \dodoi{10.1093/mnras/stx2145}

\bibitem[{{Gong} {et~al.}(2021){Gong}, {Ivlev}, {Akimkin}, \&
  {Caselli}}]{Gong2021}
{Gong}, M., {Ivlev}, A.~V., {Akimkin}, V., \& {Caselli}, P. 2021, arXiv
  e-prints, arXiv:2106.09525.
\newblock \doarXiv{2106.09525}

\bibitem[{Jiang {et~al.}(2018)Jiang, Ma, \& Ramachandran}]{Jiang2018}
Jiang, S.-Y., Ma, A., \& Ramachandran, S. 2018, International Journal of
  Molecular Sciences, 19, \dodoi{10.3390/ijms19102966}

\bibitem[{{Johansen} \& {Okuzumi}(2018)}]{Johansen2018}
{Johansen}, A., \& {Okuzumi}, S. 2018, \aap, 609, A31,
  \dodoi{10.1051/0004-6361/201630047}

\bibitem[{Jungmann {et~al.}(2022)Jungmann, Onyeagusi, Teiser, \&
  Wurm}]{Jungmann2022x}
Jungmann, F., Onyeagusi, F.~C., Teiser, J., \& Wurm, G. 2022, Journal of
  Electrostatics, 117, 103705,
  \dodoi{https://doi.org/10.1016/j.elstat.2022.103705}

\bibitem[{Jungmann {et~al.}(2021)Jungmann, van Unen, Teiser, \&
  Wurm}]{Jungmann2021b}
Jungmann, F., van Unen, H., Teiser, J., \& Wurm, G. 2021, Physical Review E,
  104, L022601

\bibitem[{{Kelling} {et~al.}(2014){Kelling}, {Wurm}, \&
  {K{\"o}ster}}]{Kelling2014}
{Kelling}, T., {Wurm}, G., \& {K{\"o}ster}, M. 2014, \apj, 783, 111,
  \dodoi{10.1088/0004-637X/783/2/111}

\bibitem[{{Krauss} {et~al.}(2003){Krauss}, {Hor{\'a}nyi}, \&
  {Robertson}}]{Krauss2003}
{Krauss}, C.~E., {Hor{\'a}nyi}, M., \& {Robertson}, S. 2003, New Journal of
  Physics, 5, 70, \dodoi{10.1088/1367-2630/5/1/370}

\bibitem[{Kruss {et~al.}(2016)Kruss, Demirci, Koester, Kelling, \&
  Wurm}]{Kruss2016}
Kruss, M., Demirci, T., Koester, M., Kelling, T., \& Wurm, G. 2016, ApJ, 827,
  110, \dodoi{10.3847/0004-637X/827/2/110}

\bibitem[{{Lee} {et~al.}(2018){Lee}, {James}, {Waitukaitis}, \&
  {Jaeger}}]{Lee2018}
{Lee}, V., {James}, N.~M., {Waitukaitis}, S.~R., \& {Jaeger}, H.~M. 2018,
  Physical Review Materials, 2, 035602,
  \dodoi{10.1103/PhysRevMaterials.2.035602}

\bibitem[{{Matsuyama}(2018)}]{Matsuyama2018}
{Matsuyama}, T. 2018, in American Institute of Physics Conference Series, Vol.
  1927, The 1st International Conference and Exhibition on Powder Technology
  Indonesia (ICePTi) 2017, 020001, \dodoi{10.1063/1.5021189}

\bibitem[{{M{\'e}ndez Harper} \& {Dufek}(2016)}]{Harper2016}
{M{\'e}ndez Harper}, J., \& {Dufek}, J. 2016, Journal of Geophysical Research
  (Atmospheres), 121, 8209, \dodoi{10.1002/2015JD024275}

\bibitem[{{Mori} {et~al.}(2021){Mori}, {Okuzumi}, {Kunitomo}, \&
  {Bai}}]{Mori2021}
{Mori}, S., {Okuzumi}, S., {Kunitomo}, M., \& {Bai}, X.-N. 2021, \apj, 916, 72,
  \dodoi{10.3847/1538-4357/ac06a9}

\bibitem[{{Ormel} \& {Cuzzi}(2007)}]{Ormel2007}
{Ormel}, C.~W., \& {Cuzzi}, J.~N. 2007, \aap, 466, 413,
  \dodoi{10.1051/0004-6361:20066899}

\bibitem[{{Schoenau} {et~al.}(2021){Schoenau}, {Steinpilz}, {Teiser}, \&
  {Wurm}}]{Schoenau2021}
{Schoenau}, L., {Steinpilz}, T., {Teiser}, J., \& {Wurm}, G. 2021, Gran. Mat.,
  23, 1

\bibitem[{{Shaw}(1917)}]{Shaw1917}
{Shaw}, P.~E. 1917, Proceedings of the Royal Society of London Series A, 94,
  16, \dodoi{10.1098/rspa.1917.0046}

\bibitem[{Skalný {et~al.}(2006)Skalný, Hortváth, \& Mason}]{Skalny2006}
Skalný, J., Hortváth, G., \& Mason, N.~J. 2006, AIP Conference Proceedings,
  876, 284, \dodoi{10.1063/1.2406037}

\bibitem[{Steinpilz {et~al.}(2019)Steinpilz, Teiser, \& Wurm}]{Steinpilz2019}
Steinpilz, T., Teiser, J., \& Wurm, G. 2019, The Astrophysical Journal, 874,
  60, \dodoi{10.3847/1538-4357/ab07bb}

\bibitem[{{Turner} {et~al.}(2014){Turner}, {Fromang}, {Gammie}, {Klahr},
  {Lesur}, {Wardle}, \& {Bai}}]{Turner2014}
{Turner}, N.~J., {Fromang}, S., {Gammie}, C., {et~al.} 2014, in Protostars and
  Planets VI, ed. H.~{Beuther}, R.~S. {Klessen}, C.~P. {Dullemond}, \&
  T.~{Henning}, 411, \dodoi{10.2458/azu\_uapress\_9780816531240-ch018}

\bibitem[{{Umebayashi} \& {Nakano}(2009)}]{Umebayashi2009}
{Umebayashi}, T., \& {Nakano}, T. 2009, \apj, 690, 69,
  \dodoi{10.1088/0004-637X/690/1/69}

\bibitem[{{Wood}(2000)}]{Wood2000}
{Wood}, J.~A. 2000, \ssr, 92, 87, \dodoi{10.1023/A:1005249417716}

\bibitem[{{Wurm} {et~al.}(2022){Wurm}, {Jungmann}, \& {Teiser}}]{Wurm2022}
{Wurm}, G., {Jungmann}, F., \& {Teiser}, J. 2022, \mnras, 517, L65,
  \dodoi{10.1093/mnrasl/slac077}

\bibitem[{{Wurm} {et~al.}(2019){Wurm}, {Schmidt}, {Steinpilz}, {Boden}, \&
  {Teiser}}]{Wurm2019}
{Wurm}, G., {Schmidt}, L., {Steinpilz}, T., {Boden}, L., \& {Teiser}, J. 2019,
  Icarus, 331, 103, \dodoi{10.1016/j.icarus.2019.05.004}

\bibitem[{Wurm \& Teiser(2021)}]{Wurm2021nat}
Wurm, G., \& Teiser, J. 2021, Nature Reviews Physics, 3, 405,
  \dodoi{10.1038/s42254-021-00312-7}

\bibitem[{{Yang} {et~al.}(2018){Yang}, {Mac Low}, \& {Johansen}}]{Yang2018}
{Yang}, C.-C., {Mac Low}, M.-M., \& {Johansen}, A. 2018, \apj, 868, 27,
  \dodoi{10.3847/1538-4357/aae7d4}

\bibitem[{{Zsom} {et~al.}(2010){Zsom}, {Ormel}, {G{\"u}ttler}, {Blum}, \&
  {Dullemond}}]{Zsom2010}
{Zsom}, A., {Ormel}, C.~W., {G{\"u}ttler}, C., {Blum}, J., \& {Dullemond},
  C.~P. 2010, Astronomy \& Astrophysics, 513, A57,
  \dodoi{10.1051/0004-6361/200912976}

\end{thebibliography}
\bibliographystyle{aasjournal}

\end{document}